\documentclass[structabstract]{aa}  
\usepackage{amssymb}
\usepackage[usenames, dvipsnames]{color}
\usepackage{graphicx}
\usepackage{natbib}
\usepackage{txfonts}
\usepackage{url}

\usepackage{hyperref}
\hypersetup{
    final=true,
    pageanchor=true,
    colorlinks=true,
    breaklinks=true,
    linkcolor=blue,
    citecolor=blue,
    urlcolor=blue,
    pdfpagemode=UseNone,
    pdftitle={Horizontal flow fields observed in Hinode G-band images.
        II. Flow Fields in the Final Stages of Sunspot Decay},
    pdfauthor={M. Verma and C. Denker},
    pdfsubject={Solar Physics},
    pdfkeywords={Sun: chromosphere, Sun: photosphere, Sun: surface magnetism,
    Sun: sunspots, Techniques: image processing, Methods: data analysis}}

\begin{document}


\title{Horizontal flow fields observed in\\
    \textit{Hinode} G-band images}
\subtitle{II.\ Flow fields in the final stages of sunspot decay}
\author{M.\ Verma\inst{1}, H.\ Balthasar\inst{1}, N.\ Deng\inst{2,3},
    C.\ Liu\inst{2}, T.\
    Shimizu\inst{4}, H.\ Wang\inst{2}, \and C.\ Denker\inst{1}}
\authorrunning{Verma et al.}
\institute{
    \inst{1} Leibniz-Institut f\"ur Astrophysik Potsdam (AIP),
    An der Sternwarte 16,
    14482 Potsdam,
    Germany\\
    \email{mverma@aip.de, hbalthasar@aip.de, cdenker@aip.de}\\
    \inst{2} New Jersey Institute of Technology, 
    Space Weather Research Laboratory, 
    323 Martin Luther King Blvd., Newark,
    NJ 07102, USA \\
    \email{nd7@njit.edu, chang.liu@njit.edu, haimin.wang@njit.edu}\\
    \inst{3} California State University Northridge,
    Physics and Astronomy Department,
    18111 Nordhoff St.,
    Northridge, CA 91330, USA\\
    \inst{4} Institute of Space and Astronautical Science,
    Japan Aerospace Exploration Agency,
    3-1-1 Yoshinodai, Chuo-ku, Sagamihara, Kanagawa 252-5210, Japan\\
    \email{shimizu.toshifumi@isas.jaxa.jp}\\}
\date{Received August 08, 2011; accepted December 02, 2011}

\abstract
{Generation and dissipation of magnetic fields is a fundamental physical process
on the Sun. In comparison to flux emergence and the initial stages of sunspot
formation, the demise of sunspots still lacks a comprehensive description. } 
{The evolution of sunspots is most commonly discussed in terms of their
intensity and magnetic field. Here, we present additional information regarding
the three-dimensional flow field in the vicinity of sunspots towards the end of
their existence.}
{We present a subset of multi-wavelengths observations obtained with the
Japanese \textit{Hinode} mission, the \textit{Solar Dynamics Observatory} (SDO),
and the \textit{Vacuum Tower Telescope} (VTT) at \textit{Observatorio del
Teide}, Tenerife, Spain during the time period from 2010 November 18--23.
Horizontal proper motions were derived from G-band and Ca\,\textsc{ii}\,H
images, whereas line-of-sight velocities were extracted from VTT Echelle
H$\alpha$~$\lambda656.28$~nm spectra and Fe\,\textsc{i} $\lambda630.25$~nm
spectral data of the \textit{Hinode/Spectro-Polarimeter}, which also provided
three-dimensional magnetic field information. The \textit{Helioseismic and
Magnetic Imager} on board SDO provided continuum images and line-of-sight
magnetograms as context for the high-resolution observations for the entire disk
passage of the active region.}
{We have performed a quantitative study of photospheric and chromospheric flow
fields in and around decaying sunspots. In one of the trailing sunspots of
active region NOAA~11126, we observed moat flow and moving magnetic features
(MMFs), even after its penumbra had decayed. We also noticed a superpenumbral
structure around this pore. MMFs follow well-defined,
radial paths from the spot all the way to the border of a supergranular cell
surrounding the spot. In contrast, flux emergence near the other sunspot
prevented it from establishing such well ordered flow patterns, which could even
be observed around a tiny pore of just 2~Mm diameter. After the disappearance of
the sunspots/pores a coherent patch of abnormal granulation remained at their
location, which was characterized by more uniform horizontal proper motions, low
divergence values, and diminished photospheric Doppler velocities. This region,
thus, differs significantly from granulation and other areas covered by G-band
bright points. We conclude that this peculiar flow pattern is a signature of
sunspot decay and the dispersal of magnetic flux.}
{}


\maketitle


\begin{figure*}[t]
\includegraphics[width=\textwidth]{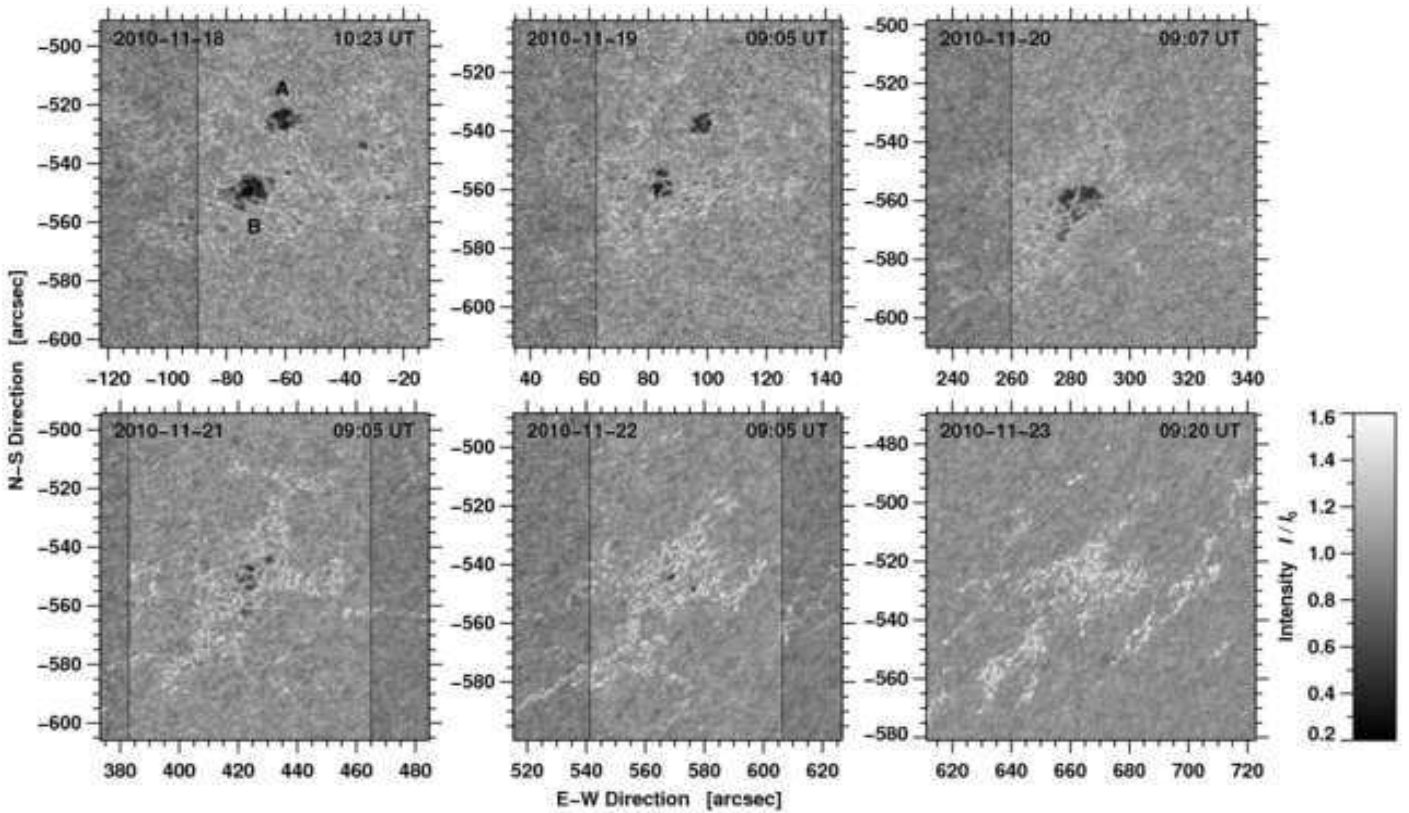}
\caption{Calibrated \textit{Hinode} G-band images showing the decay of active
    region NOAA~11126 observed during the period from 2010 November 18--23
    (\textit{from top-left to bottom-right}). The FOV is $111\arcsec \times
    111\arcsec$. The annotation of the axes refers to heliocentric coordinates
    given in seconds of arc. Brighter areas delineated by vertical black lines
    correspond to regions, which were covered by spectral scans obtained with
    the VTT Echelle spectrograph. The intensity scale to the right applies to
    these regions, while areas not covered by Echelle data are displayed with an
    offset of $0.1 I_0$. On 2010 November~23, only few \textit{Hinode} G-band
    images were available. The limb darkening has been subtracted from the
    G-band
    images, which were then normalized so that the mean of the quiet Sun
    intensity distribution corresponds to unity.}
\label{FIG01}
\end{figure*}

\section{Introduction}

Sunspots are a thought-provoking aspect of solar activity because of the close
interaction between plasma motions and magnetic fields. Recent progress in MHD
simulations \citep[e.g.,][]{Rempel2011} provide a comprehensive framework for
the interpretation of high-resolution sunspot observations. The formation of a
penumbra around a sunspot is a rapid phenomenon, i.e., within a few hours a
sunspot can develop a penumbra \citep{Leka1998, Yang2003a}, which is intimately
linked to more inclined magnetic field lines and the onset of the Evershed flow.
\citet{Schlichenmaier2010b} observed the growth of a penumbra where only the
newly formed penumbra contributed to the increase in spot size while the umbra
remained stable. The formation of a penumbra, which would surround the entire
spot, was hindered by continuous flux emergence between the spots of the bipolar
region.

Quite the opposite, the decay of a sunspot is a slow process. Decay rates for
stable leading sunspots and irregular follower spots are different
\citep{MartinezPillet2002}. A number of decay laws were proposed such as a
linear decay law described by \citet{Bumba1963} and a parabolic decay law
proposed by \citet{Petrovay1997}. \citet{MartinezPillet2002} critically reviews
various diffusion models, concludes that they explain well how flux is spread
over larger areas while the spot is decaying, but they fail to satisfactorily
describe the flux removal process. The initial stages of sunspot decay, i.e.,
when the spot looses its penumbra, are exemplary described in
\citet{BellotRubio2008}, who discovered finger-like structures, which are
neither related to penumbral filaments nor the Evershed flow. These features
might be penumbral field lines rising to the chromosphere, thus contributing to
the decay of the sunspot penumbra. When a sunspot looses its penumbra, its decay
reaches a critical point. Magnetic field lines become more vertical and
convective motions in its vicinity begin to change. These ideas of a critical
inclination angle and convective motions were put together by
\citet{Rucklidge1995}, who explains in a simple model why small sunspots can
have a penumbra while larger pores do not possess one.

The moat flow is a large-scale flow pattern commonly observed around sunspots
\citep{Meyer1974}. However, flux removal and dispersal can only be understood in
the context of the moat flow's fine structure. Moving magnetic features (MMFs)
play a major role in the flux dispersal process and they are only associated
with decaying sunspots \citep{Harvey1973}. The total flux carried by MMFs is
several times larger than the flux contained within the sunspot itself. Thus,
the polarity of MMFs has to be considered for a balance of the net flux. MMFs
move radially outward with a velocity of 1~km~s$^{-1}$ before they reach and
dissolve within the network, i.e., at the boundaries of the supergranular cell
containing the sunspot. \citet{Zuccarello2009} showed evidence that MMFs and
moat flow  are present even in the vicinity of pores, i.e., in the absence of
penumbral filaments and Evershed flow \citep[cf.,][]{CabreraSolana2006}.
\citet{Deng2007} also detected a persistent moat flow after the penumbra around
spot disappeared leaving only a pore. Even though the moat flow might not be
closely tied to the Evershed flow, the sub-photospheric interaction of magnetic
field lines and flows can still produce the observed flow patterns.

\citet{Verma2011} described a local correlation tracking (LCT) method to measure
horizontal flows based on \textit{Hinode} G-band images. In this study, we
perform a case study, where we put such horizontal flow fields in the context of
other photospheric and chromospheric data. In particular, we are interested in
the final stages of sunspot decay. In Sect.~\ref{SEC02}, we present a subset of
multi-wavelengths observations, which were obtained within the scope of
\textit{Hinode Operation Plan} (HOP) 0176. The temporal evolution of active
region NOAA~11126 in terms of intensity, morphology, and flow as well as
magnetic fields is described in Sect.~\ref{SEC03} and discussed in
Sect.~\ref{SEC04}.

\begin{table*}[t]
\caption{Observing characteristics and physical parameters}\label{TAB01}
\tiny
\begin{tabular}{llcccccp{65mm}}
\cline{1-7}\vspace*{-2.5mm}\\\cline{1-7}
\multicolumn{2}{l}{November} & 18 & 19 & 20 & 21 & 22 & \rule[-3pt]{0pt}{12pt}\\
\cline{1-7}
$B$                  & & S32.6$^{\circ}$ & S32.6$^{\circ}$ & S32.6$^{\circ}$ & S32.6$^{\circ}$ & S32.6$^{\circ}$ & \rule{0pt}{9pt}\\
$L$                  & & E5.5$^{\circ}$  & W6.3$^{\circ}$  & W20.5$^{\circ}$ & W31.5$^{\circ}$ & W44.1$^{\circ}$ & \\
$\mu$                &  & $0.81$  & $0.81$ &  $0.77$ & $0.70$ & $0.58$ & \\
$t_{0,\mathrm{GB}}$  & UT & 10:23 & 09:05 & 09:07 & 09:05 & 09:05 &  \\
$t_{0,\mathrm{SP}}$  & UT & 11:46 & 09:05 & 09:07 & 09:04 & 09:04 &  \\
$t_{0,\mathrm{ES}}$  & UT & 10:22 & 08:59 & 14:16 & 09:31 & 08:47 &  \rule[-4pt]{0pt}{6pt}\\
\cline{1-7}
$A_\mathrm{Spot\ A}$ & Mm$^{2}$          & 33.8               &  21.5           &     &     &     & \rule{0pt}{9pt}\\
$v_\mathrm{Spot\ A}$ & km s$^{-1}$       & $0.22 \pm 0.19$    & $0.15 \pm 0.15$ &     &     &     & \\
                     &                   & $0.20 \pm 0.25$    & $0.15 \pm 0.14$ &     &     &     & \\
$v_\mathrm{Ring\ A}$ & km s$^{-1}$       & $0.44 \pm 0.23$    & $0.37 \pm 0.19$ &     &     &     & \\
                     & km s$^{-1}$       & $0.44 \pm 0.23$    & $0.38 \pm 0.19$ &     &     &     & \\         
$\left|\nabla v\right|_\mathrm{Ring\ A}$ & $10^{-3}$ s$^{-1}$ & $0.83 \pm 0.78$ & $0.65 \pm 0.54$ &      &     &     & \\
                                         &                    & $1.28 \pm 1.33$ & $1.13 \pm 1.06$ &      &     &     & \rule[-4pt]{0pt}{6pt}\\  
\cline{1-7}
$A_\mathrm{Spot\ B}$ & Mm$^{2}$          & 54.7               & 21.6            & 54.2            & 17.5            & 2.7 & \rule{0pt}{9pt}\\
$v_\mathrm{Spot\ B}$ & km s$^{-1}$       & $0.22 \pm 0.15$    & $0.15 \pm 0.15$ & $0.16 \pm 0.07$ & $0.18 \pm 0.05$ & $0.28 \pm 0.06$ & \\
                     &                   & $0.20 \pm 0.14$    & $0.11 \pm 0.05$ & $0.15 \pm 0.08$ & $0.25 \pm 0.06$ & $0.37 \pm 0.06$ & \\
$v_\mathrm{Ring\ B}$ & km s$^{-1}$       & $0.31 \pm 0.19$    & $0.20 \pm 0.12$ & $0.28 \pm 0.14$ & $0.25 \pm 0.14$ & $0.30 \pm 0.09$ & \\
                     &                   & $0.32 \pm 0.20$    & $0.20 \pm 0.12$ & $0.29 \pm 0.15$ & $0.28 \pm 0.13$ & $0.40 \pm 0.09$ & \\
$\left|\nabla v\right|_\mathrm{Ring\ B}$ & $10^{-3}$ s$^{-1}$ & $0.63 \pm 0.68$ & $0.37 \pm 0.36$ & $0.46 \pm 0.49$ & $0.47 \pm 0.51$ & $0.28 \pm 0.33$ & \\
                                         &                    & $0.91 \pm 1.20$ & $0.47 \pm 0.59$ & $0.76 \pm 1.04$ & $0.63 \pm 0.81$ & $0.44 \pm 0.44$ &  \rule[-4pt]{0pt}{6pt}\\
\cline{1-7}
$A_\mathrm{mag}$  &  Mm$^{2}$   & $136.9$ & $83.5$& $98.9$ & $44.7$ & $22.2$ & \rule{0pt}{9pt}\\
$N_\mathrm{mag}$  &             & 10      & 5     & 9      & 6      & 2 & \\
$v_\mathrm{mag}$  & km s$^{-1}$ & $0.22 \pm 0.12$ & $0.15 \pm 0.09$ & $0.16 \pm 0.07$ & $0.18 \pm 0.05$ & $0.28 \pm 0.06$ & \\
$v_\mathrm{bp}$   & km s$^{-1}$ & $0.27 \pm 0.19$ & $0.27 \pm 0.16$ & $0.28 \pm 0.16$ & $0.23 \pm 0.12$ & $0.26 \pm 0.10$ & \\
$v_\mathrm{gran}$ & km s$^{-1}$ & $0.40 \pm 0.24$ & $0.35 \pm 0.20$ & $0.37 \pm 0.20$ & $0.34 \pm 0.19$ & $0.34 \pm 0.18$ & \\
$\left|\nabla v\right|_\mathrm{gran}$ & $10^{-3}$ s$^{-1}$ & $0.88 \pm 0.89$ & $0.72 \pm 0.67$ & $0.85 \pm 0.79 $ & $0.95 \pm 0.88$ & $1.04\pm 0.98$ & \\
$\left|v\right|_\mathrm{mag, LOS}$    & km s$^{-1}$        & $0.03 \pm 0.17$ & $0.02 \pm 0.12$ & $0.04 \pm 0.20$  & $0.02 \pm 0.12$ & $0.00 \pm 0.02$ &\\
$\left|v\right|_\mathrm{bp, LOS}$     & km s$^{-1}$        & $0.43 \pm 0.37$ & $0.44 \pm 0.37$ & $0.44 \pm 0.38$  & $0.41 \pm 0.38$ & $0.34 \pm 0.30$ &\\
$\left|v\right|_\mathrm{gran, LOS}$   & km s$^{-1}$        & $0.51 \pm 0.41$ & $0.50 \pm 0.41$ & $0.55 \pm 0.45$  & $0.57 \pm 0.47$ & $0.55 \pm 0.45$ &\rule[-4pt]{0pt}{6pt}
\raisebox{23mm}[-23mm]{\parbox{65mm}{
The parameters in the first column of the table refer to
     heliographic latitude $B$,
     heliographic longitude $L$,
     cosine of the heliocentric angle $\mu$,
     start of the G-band (GB),
     spectro-polarimeter (SP),
     and Echelle spectrograph (ES)
     observing sequences $t_{0}$,
     spot area $A$,
     horizontal velocity $v$,
     mean divergence $\left|\nabla v\right|$,
     number of magnetic elements $N_\mathrm{mag}$.
The indices refer to G-band bright points (bp), granulation (gran), magnetic
elements (mag), the two sunspots (Spot~A and Spot~B), and the four-megameter
wide annuli around both spots (Ring~A and Ring~B). If two rows are given for a
physical parameter, then the top row refers to G-band data, whereas the bottom
row was derived from Ca\,\textsc{ii}\,H data. If present, the standard deviation
refers to the variation of the physical parameters  within the specified regions
rather than to any formal error.}}\\
\cline{1-7}
\end{tabular}
\end{table*}


\section{Observations\label{SEC02}}

The disk passage of active region NOAA~11126 started on 2010 November~12 and
ended on November~24. NOAA~11126 was classified as a $\beta$-region, while it
crossed the solar disk. No major flaring was associated with the region. Only a
few B-class events were reported on 2010 November 15. As part of HOP~0176
``\textit{High-resolution multi-wavelength study of small-scale jets on the
solar disk}'', we observed the decay of two small follower sunspots in the
active region for five days from 2010 November 18--22. A time-series of
\textit{Hinode} G-band images is shown in Fig.~\ref{FIG01}, where we labeled the
northern and southern spots with \textsf{A} and \textsf{B}, respectively. In the
following, we will simply refer to these magnetic features as \textit{spots},
even if the proper classification should be \textit{pores}, i.e., sunspots
lacking a penumbra. Spectral scans with different field-of-views (FOVs) and
cadences were observed for three hours every day. Since we are only focusing on
the general properties of sunspot decay, we chose the first scan on the given
day, which covered the largest FOV. The settings for the \textit{Vacuum Tower
Telescope} (VTT) Echelle data and data of \textit{Hindoe/Spectro-Polarimeter-}
were chosen as to obtain the best spatial and spectral match. The general
observing characteristics are listed in Tab.~\ref{TAB01}.

\subsection{SDO/HMI full-disk images\label{SEC02.1}}

The discussion of the temporal evolution and morphology of active region
NOAA~11126 is based on full-disk images and line-of-sight (LOS) magnetograms
obtained with the \textit{Helioseismic and Magnetic Imager}
\citep[HMI,][]{Schou2010, Couvidat2011, Wachter2011} on board the \textit{Solar
Dynamics Observatory} (SDO). Since the \textit{Hinode} FOV is too small to cover
the entire active region and to give an overview of the magnetic field topology,
we show in Fig.~\ref{FIG02} the limb-darkening corrected HMI continuum image and
magnetogram for November~18.

We selected from the SDO/HMI database one image/magnetogram with $4096 \times
4096$ pixels every 15~min for the period from 2010 November 13--23, i.e., a
total of 1056 full-disk images. The image scale is about 0.5\arcsec\
pixel$^{-1}$, so that even finer details of penumbra and umbra can be captured.
The average limb-darkening function was computed for this time interval and
subtracted from the full-disk images to yield contrast enhanced images
\citep[see e.g.,][]{Denker1999a}, which can then be used for feature
identification. The photometric temporal evolution for the
entire active region as shown in Fig.~\ref{FIG02} is depicted in
Fig.~\ref{FIG03}. The corresponding changes of the magnetic flux are shown in
Fig.~\ref{FIG04}. Since HMI data cover the whole solar disk, it is
straightforward to compute the heliocentric angle $\mu$ on a pixel-by-pixel
basis. Thus, a geometrical correction is applied to the measured areas and flux
values of Figs.~\ref{FIG03} and \ref{FIG04}, which are discussed in detail in
Sects.~\ref{SEC03.1} and \ref{SEC03.5}, respectively.

\subsection{Hinode G-band and \ion{Ca}{ii}\ H images\label{SEC02.2}}

We applied LCT \citep[for details see][]{Verma2011} to image sequences captured
in G-band $\lambda$430.5~nm and \ion{Ca}{ii}\ H {$\lambda$396.8~nm} to compare
horizontal flows in the  photosphere and chromosphere. Note that the
\ion{Ca}{ii}\ H images do not purely represent the chromosphere, but contain
contributions from both the upper photosphere and lower chromosphere. These
observations were carried out by the \textit{Broad-band Filter Imager} (BFI) of
the \textit{Solar Optical Telescope} \citep[SOT,][]{Tsuneta2008} on board
\textit{Hinode} \citep{Kosugi2007}. Data sequences were captured every day from
09:00~UT to 12:00~UT with an average time cadence of 120~s (with some jumps in
the data sequences). In both wavelengths the images are $2 \times 2$-pixel
binned with an image scale of 0.11\arcsec\ pixel$^{-1}$. Images have a size of
$1024 \times 1024$ pixels and a FOV of $111\arcsec \times 111\arcsec$.

After basic data calibration, the images were corrected for geometrical
foreshortening and resampled onto a regular grid of $80\ \mathrm{km}\times 80\
\mathrm{km}$. The signature of the five-minute oscillation was removed from the
images by using a three-dimensional Fourier filter with a cut-off velocity of
8~km~s$^{-1}$ corresponding roughly to the photospheric sound speed. For
measuring horizontal proper motions, we applied the LCT algorithm described in
\citet{Verma2011}, which computes cross-correlations over $32 \times 32$-pixel
regions with a Gaussian kernel having a FWHM of 15 pixels (1200~km)
corresponding to the typical size of a granule. In two
aspects we deviated from the aforementioned algorithm,
the time cadence was $\Delta t = 120$~s and the flow maps were averaged over
$\Delta T=3$~h.

\subsection{H$\alpha$ Echelle spectra}

The observations in H$\alpha$ $\lambda$656.28~nm and Fe\,\textsc{i}
$\lambda$656.92~nm were carried out with the VTT Echelle spectrograph. Spectral
data were acquired with a slit width of 80~$\mu$m and an exposure time of
300~ms. The image-scale of the spectrograph is
$8.99\arcsec$ mm$^{-1}$. We did not use a predisperser, hence, to suppress
overlapping in spectral orders, we placed a broad-band interference filter
directly behind the spectrograph slit. The infrared grating with a blaze angle
of $51.6^\circ$ and 200 grooves~mm$^{-1}$ was used to record spectra in the
$12^\mathrm{th}$ order. In this configuration, we achieved a dispersion of
$0.60$~pm~pixel$^{-1}$. The spectra covered a wavelength range of 1.2~nm from
$\lambda 655.9$~nm to $\lambda 657.1$~nm. We employed a PCO.4000 CCD camera that
has a quantum efficiency of about 30\% at H$\alpha$. After $2 \times 2$-pixel
binning the images have a size of $2004 \times 1336$ pixels. The pixel size of
the CCD detector is 9~$\mu$m $\times$ 9~$\mu$m. The Echelle  data were intended
to match the \textit{Hinode} observations, which was mostly achieved except for
a few interruptions because of deteriorating seeing conditions. The
two-dimensional FOV was scanned with a spatial step of 0.32\arcsec\ and 200--250
spectra were recorded in a sequence. A sequence of 240 spatial steps took about
12 min and covered a FOV of 72.0\arcsec $\times$ 182.6\arcsec.

\begin{figure}[t]
\centerline{\includegraphics[width=\columnwidth]{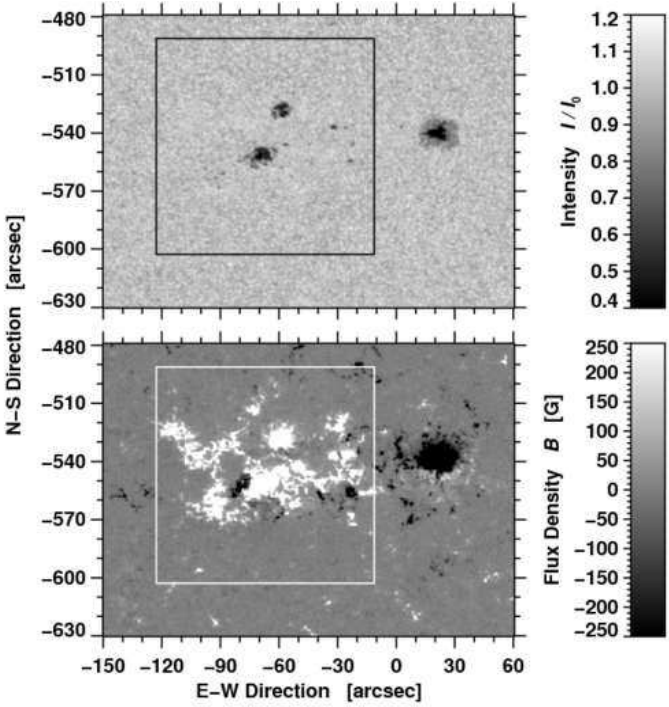}}
\caption{Limb-darkening corrected HMI continuum image (\textit{top}) and
    magnetogram (\textit{bottom}) of active region NOAA~11126 on November~18.
    The square box in both images shows the FOV covered by \textit{Hinode}/BFI.
    The axes are labeled in heliographic coordinates.}
\label{FIG02}
\end{figure}

The common FOVs of the G-band images and H$\alpha$ Echelle spectra are shown in
Fig.~\ref{FIG01} for each observing day. We first matched the image scale of the
Echelle data to that of the \textit{Hinode} data. We then aligned
\textit{Hinode} G-band images and continuum images derived from Echelle spectra.
After this procedure, the heliocentric coordinates for G-band images and
H$\alpha$ spectra differ by less than 1\arcsec\ in the periphery of the FOV.
Furthermore, the Echelle scan direction is not perfectly aligned with the
\textit{Hinode} BFI detector. Hence, we computed an offset angle (smaller than $
\pm 2^{\circ}$) for each date and applied it to the spectral data. On 2010
November~20, the time difference between G-band images and Echelle spectra was
about five hours.

\begin{figure}[t]
\centerline{\includegraphics[width=\columnwidth]{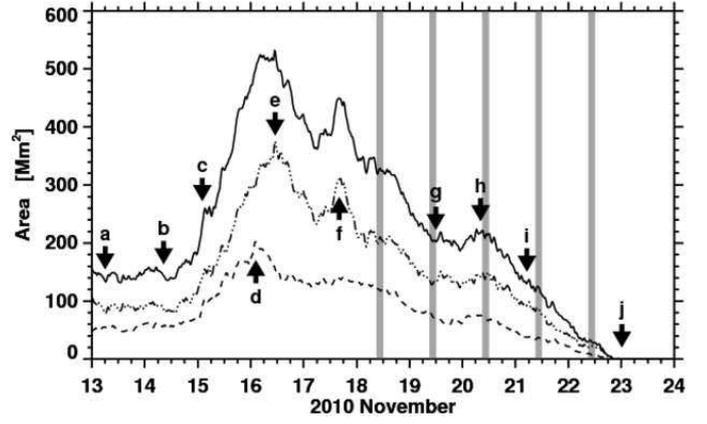}}
\caption{Temporal evolution of the area covered by active region NOAA~11126
    while it crossed the solar disk. The area enclosed by umbral cores and pores
    is displayed with dashed line. Dashed dot dot line refers to the penumbra,
    whereas solid line denotes the total area. Some smoothing was applied to the
    time-series to suppress features on temporal scales below one hour. The
    vertical gray bars refer to the observing periods of HOP~0176 (2010 November
    18--22). The labels indicate different stages of active region evolution,
    which are explained in Sect.~\ref{SEC03.1}.}
\label{FIG03}
\end{figure}

\begin{figure}[t]
\centerline{\includegraphics[width=\columnwidth]{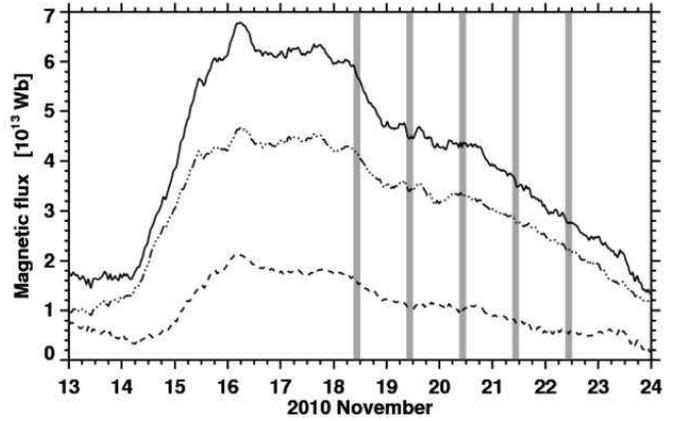}}
\caption{Temporal evolution of the magnetic flux contained in active region
   NOAA~11126 while it crossed the solar disk. The solid, dashed dot dot, and
   dashed lines refer to the total, positive, and negative flux, respectively.}
\label{FIG04}
\end{figure}

\subsection{Hinode Spectro-Polarimeter}

The photospheric magnetic topology and evolution of the active region were
studied using high-resolution spectral data from the \textit{Hinode/
Spectro-Polarimeter} \citep[SP,][]{Ichimoto2008}, which uses two magnetically
sensitive \ion{Fe}{i} lines at $\lambda 630.15$~nm and $\lambda 630.25$~nm and
the nearby continuum \citep{Tsuneta2008} to obtain Stokes $IQUV$ spectral
profiles. We used spectral data captured in the fast mapping mode with a FOV of
$58 \arcsec \times 122 \arcsec$ and an average scan time of 12~min. On
November~18, we also used ten continuous high cadence scans with a FOV of
$32\arcsec \times 123\arcsec$ and a scan time of 7~min. The dispersion is about
2.155~pm~pixel$^{-1}$. The region was scanned with a spatial step of about
0.3\arcsec\ and image scale of 0.32\arcsec\ pixel$^{-1}$. The basic data
reductions such as subtraction of dark current, flat fielding, polarization, and
wavelength correction were performed using procedures available in SolarSoft
\citep[SSW,][]{Bentley1998, Freeland1998}.


\section{Results\label{SEC03}}

\subsection{Photospheric evolution\label{SEC03.1}}

Contrast-enhanced HMI full-disk continuum images were used to study the
evolution and decay of active region NOAA~11126 during its disk passage. The
curves in Fig.~\ref{FIG03} correspond to the areas of
umbral cores/pores, penumbrae, and the sum of both types of features. These
strong magnetic features are identified according to intensity thresholds of
75\% and 92\%, respectively, where the quiet Sun intensity was normalized to
unity. Some spatial smoothing and minimum-size criteria were applied to binary
masks of identified features using morphological image processing techniques,
thus ensuring unwrinkled boundaries and contiguous structures. Note that this
algorithm only provides a rough estimate of the above areas. Some (small)
features could be misclassified. In particular, the borders of pores will be
classified as penumbra.

\begin{figure*}[t]
\includegraphics[width=\textwidth]{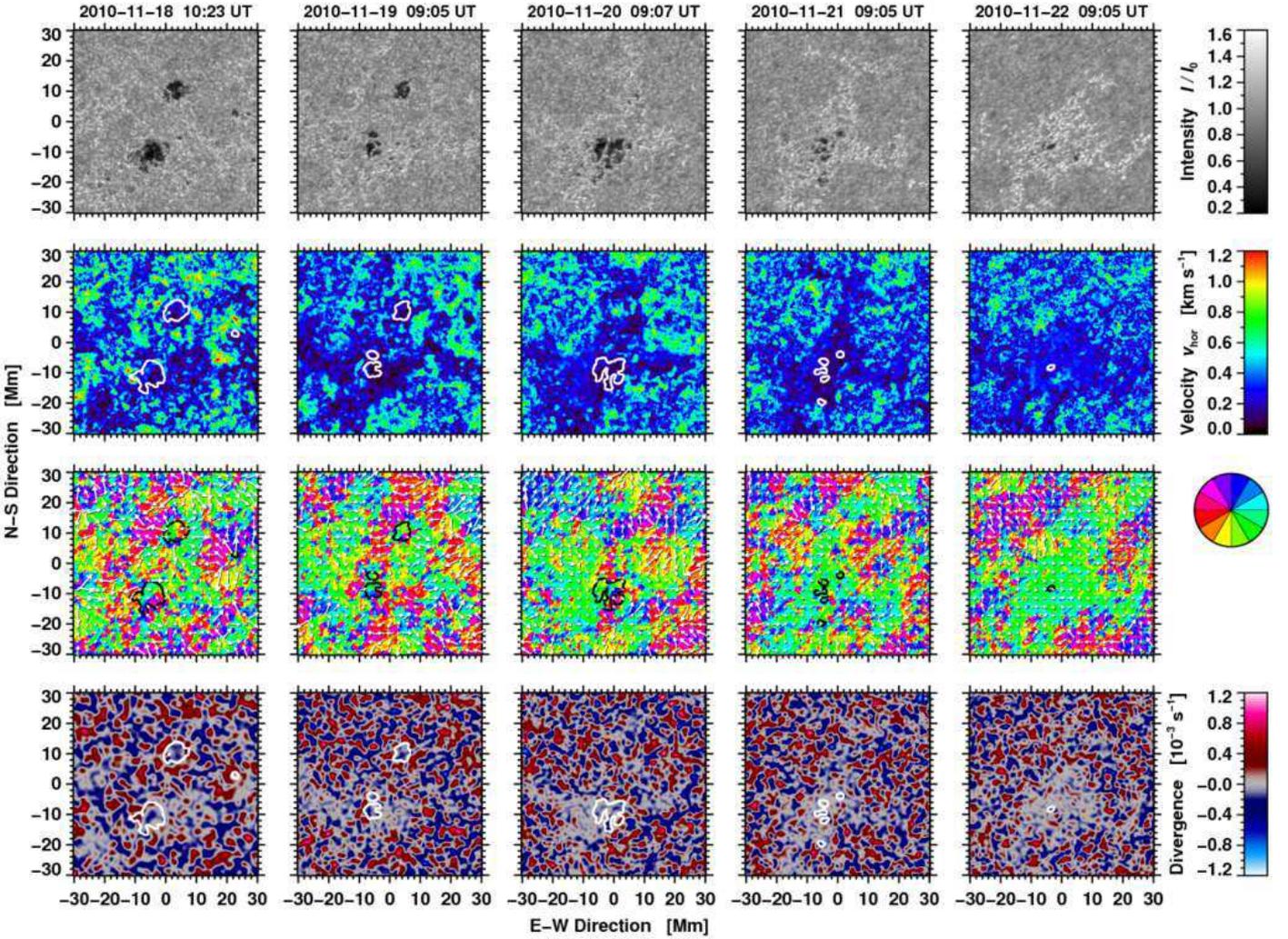}
\caption{G-band images (\textit{top}) after correction of geometrical
    foreshortening tracing the photospheric evolution of the central part
    of active region NOAA~11126 (\textit{from left to right} 2010 November
    18--22). The horizontal flow speeds (\textit{2nd row}) are given by
    the velocity scale to the right. The direction of the horizontal flows 
    (\textit{3rd row}) are displayed according to the color-wheel and
    arrows for which a velocity of 1~km~s$^{-1}$ corresponds to exactly the
    grid spacing. The divergence of the horizontal flow field (\textit{bottom})
    are presented according to the scale on the right, where gray indicates
    divergence values close to zero. All G-band images and LCT maps were aligned
    so that the center of the panels coincides with a latitude of $31.9^{\circ}$
    South. The white and black contours outline the location of the small
    sunspot/pores. Times refer to the first image
    of the time-series, which was used to compute the flow maps.}
\label{FIG05}
\end{figure*}

The most important stages of the active region evolution are labeled in
Fig.~\ref{FIG03}: (a) Initially, two tightly spaced sunspots of positive
polarity were present early on November~13. The leading spot was larger and had
a well established penumbra. (b) New flux emerged towards the south-east of
these spots at 8:30~UT on November~14. Numerous (more than ten) magnetic knots
and pores appeared to the south-west forming a bipolar magnetic region. (c) The
umbral core of leading sunspot of the new group established at 2:00~UT on
November~15. (d) The leading sunspot of the new group continuously grew by
advecting magnetic knots and small pores. The umbral cores/pores occupied the
largest area at 2:20~UT on November 16. (e) The penumbra of the leading spot
reached its maximum about nine hours later. At this time, the active region
NOAA~11126 started its decay phase. (f) Some further flux emergence occurred in
the trailing part of the region at about 16:00~UT on November~17, which
strengthened spot~\textsf{B} and produced thin elongated dark lanes. These
typical features of flux emergence \citep[see e.g.,][]{Strous1996} were labeled
erroneously as penumbrae by the thresholding algorithm.

HOP~0176 focused on the two trailing spots/pores, of which the northern one
(spot \textsf{A}) was already decaying, while the southern one (spot \textsf{B})
had just reached its maximum. This sunspot also showed strong indications of
rotation. (g) This spot then fragmented into numerous magnetic knots until about
19:30~UT on November~19. (h) At this time, the fragments started to converge
again forming a small sunspot, which reached its maximum at 8:00~UT on
November~20. Interestingly, the northern pore faded away in parallel to this
growths spurt. (i) Most of the penumbra in the leading spot has disappeared by
5:00~UT on November~21. (j) Finally, on November~23, all spots, pores, and
magnetic knots had vanished and only a bright plage region remained until it
rotated off the visible hemisphere.

In summary and neglecting all details of active region evolution, active
region decay rates can be computed using a linear fit for time periods when
the area coverage reached its maximum to the point when the area fell below
5~Mm$^{2}$. The overall decay rate of the active region is 72.6~Mm$^{2}$~per
day. The values for penumbrae and umbral cores/pores are 48.3 and
25.1~Mm$^{2}$~per day, respectively. Similarly, we computed the growth rate of
umbrae and penumbrae starting at 21:00~UT on November~14. The values are 171.4
and 104.8~Mm$^2$ per day, respectively. This is about four times faster than the
corresponding decay times but less than one half of the penumbral growth rate of
about 400~Mm$^2$ per day presented by \citet{Schlichenmaier2010b}.

\begin{figure*}[t]
\includegraphics[width=\textwidth]{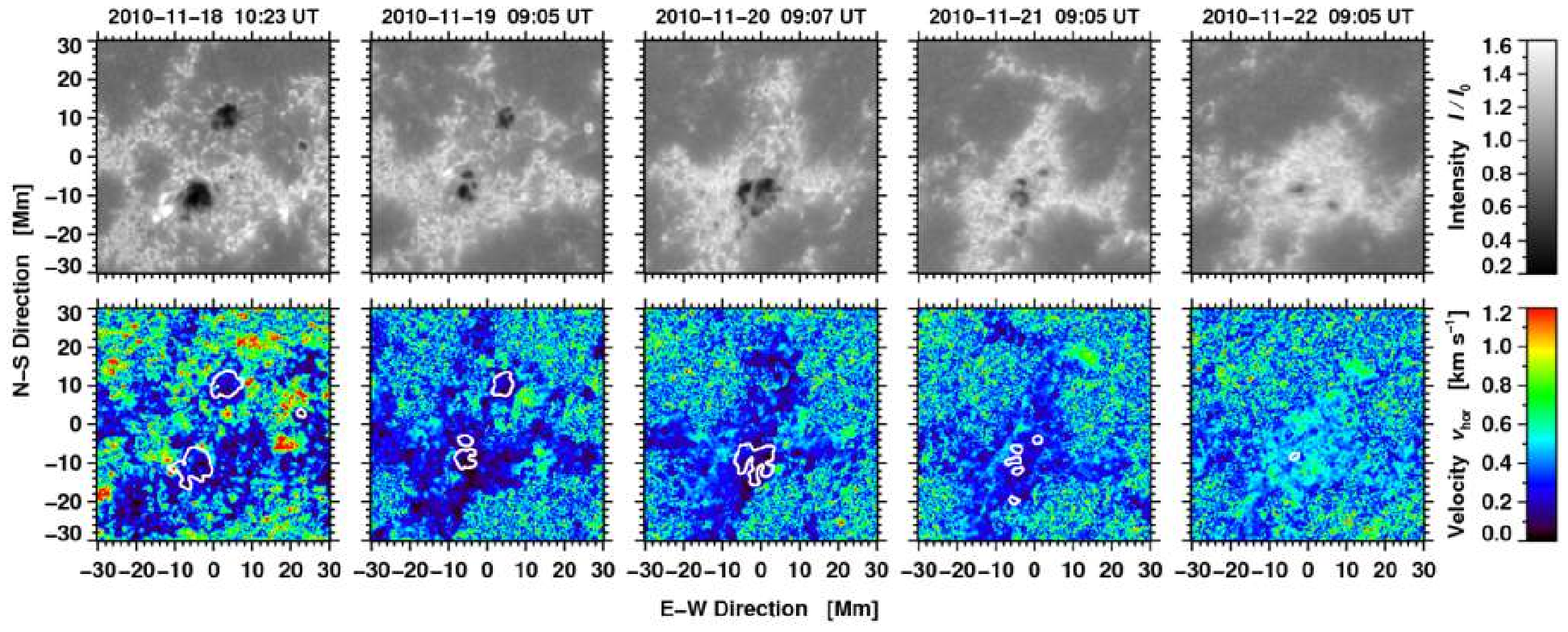}
\caption{\ion{Ca}{ii}\,H average images tracing the evolution of active region
    NOAA~11126 in the chromosphere. The horizontal proper motions were derived
    from time-sequences of \ion{Ca}{ii}\,H images. Otherwise, data processing
    and display are the same as in Fig.~\ref{FIG05}.}
\label{FIG06}
\end{figure*}

\textit{Hinode} G-band images allow us to zoom in on the two trailing spots. The
region-of-interest (ROI) is shown in the top row of Fig.~\ref{FIG05}. The data
of 2010 November 18--22 are corrected  for geometrical foreshortening and the
center-to-limb variation. The ROI with a size of $756 \times 756$ pixels or
60~Mm $\times$ 60~Mm was centered at a heliographic latitude of S$31.9^{\circ}$.
The first G-band image of the daily observing sequences was used as a reference
to align all other data.

Various solar features are identified using intensity thresholds and
morphological image processing. We indiscriminately used a fixed intensity
threshold of $I_\mathrm{mag} = 0.8 I_0$ for strong magnetic features and an
adaptive threshold for G-band bright points of $I_\mathrm{bp} = (1.37 - 0.08\mu)
I_0$, where $\mu$ is the cosine of the heliocentric angle $\theta$, and $I_0$
refers to the average quiet Sun intensity. Intensity values between
$I_\mathrm{bp}$ and $I_\mathrm{mag}$ consequently enclose granulation. The
measured spot areas for HMI and \textit{Hinode} agree with each other, and the
remaining differences can be attributed to different spectral characteristics of
the observed passbands, image scales,  and threshold/selection criteria. In
general, the temporal evolution within the ROI follows the same trend as
discussed in the context of SDO observations (Fig.~\ref{FIG03}). Areas,
velocities, and other physical quantities based on high-resolution data are
included in Tab.~\ref{TAB01}.

In the following, we present a chronology of the important phases of sunspot
decay based on the high-spatial resolution G-band images. On November ~18, the
two trailing sunspots were embedded in a network of G-band bright points hinting
at widely dispersed, weak magnetic fields. Both spots were filled with numerous
umbral dots. Spot~\textsf{A} had three umbral cores separated by faint light
bridges and a small penumbral segment pointing westward. Spot~\textsf{B} had two
umbral cores with a few associated magnetic knots. It possessed curved penumbral
filaments pointing eastwards, which are indicative of twisted magnetic field
lines. These kind of non-radial penumbral filaments are frequently observed in
flaring sunspots with horizontal shear flows \citep{Denker2007d,Deng2006}. On
November~19, the faint light-bridges had disappeared and the penumbra had
vanished leaving only a single pore, which was filled with conglomerates of
umbral dots. Spot~\textsf{B} also lost its penumbra leaving four umbral cores
separated by faint light-bridges, which, however, were more pronounced in
comparison to previous day. Magnetic knots were still surrounding
spot~\textsf{B}. Spot~\textsf{A} had decayed on November~20 with only two faint
magnetic knots remaining at its location. Interestingly, at this point of time
spot~\textsf{B} started to grow in area with hints of penumbral filament being
visible on its western side. It consisted of three umbral cores, which were now
separated by strong light-bridges, which split the spot in two halves along its
north-south axis. The presence of strong light-bridges might suggest the
initiation of the spot's decay phase  \citep[see][]{Sobotka1993}. Once
spot~\textsf{A} had disappeared on November~21, it did not leave any significant
trace within the network of G-band bright points. By this time spot~\textsf{B}
had dissolved into multiple tiny pores and magnetic knots, which roughly covered
the same region as on the previous day. By November~22, only two tiny pores were
left from spot~\textsf{B}, which disappeared on November~23 just leaving G-band
bright points at its point of disappearance. Two observations are noteworthy:
(1) The area covered by the G-band bright points remained almost constant during
the disk passage of the active region, which suggests that flux decays more or
less in place and is not redistributed over a larger area. The time scale of
flux removal or dispersal extends well beyond the photometric decay time of
strong magnetic features such as sunspots, pores, and magnetic knots. (2) The
two trailing spots have different histories of flux emergence and decay.

\begin{figure*}[t]
\sidecaption
\includegraphics[width=58mm]{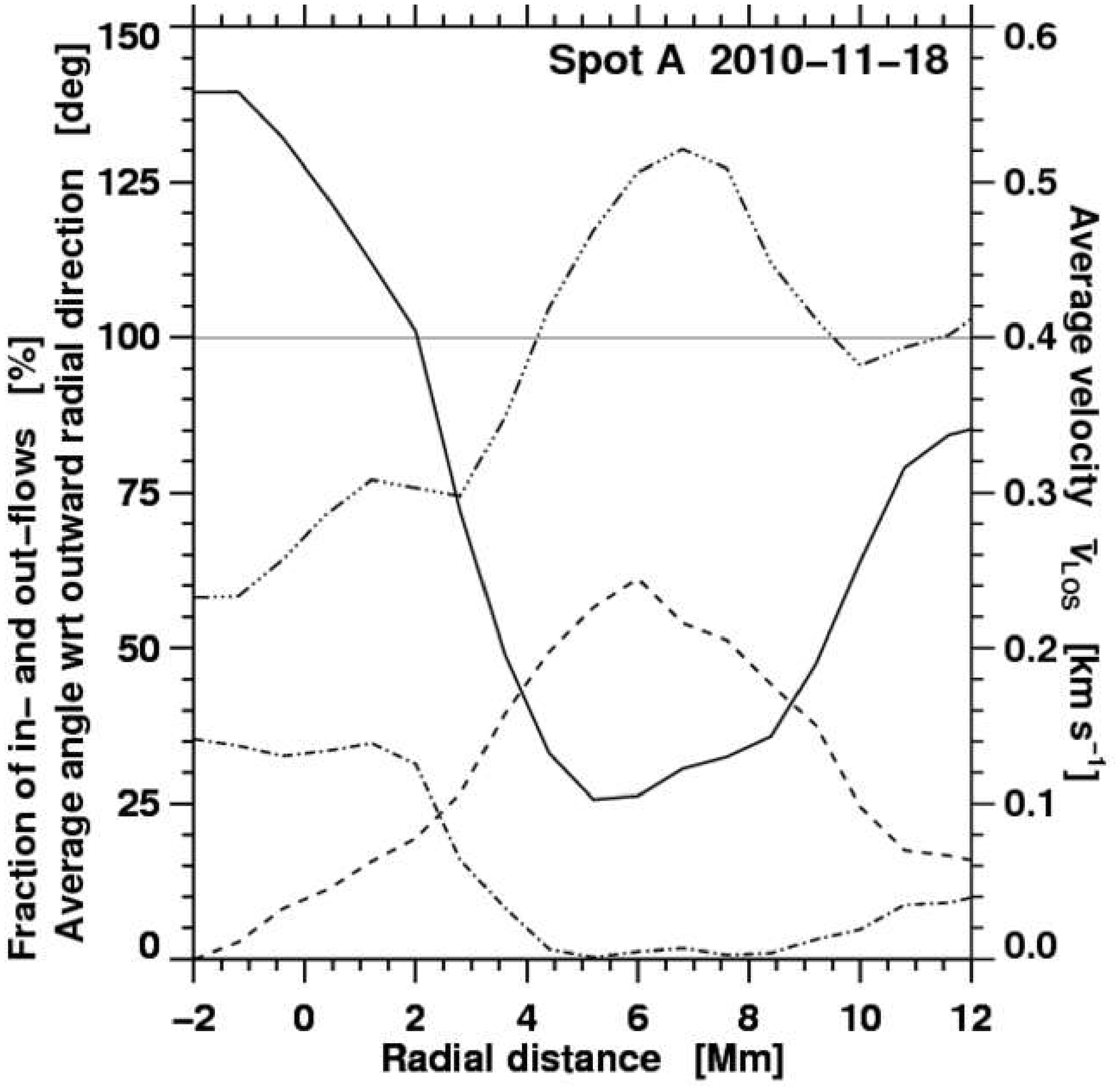}
\hspace*{4mm}
\includegraphics[width=58mm]{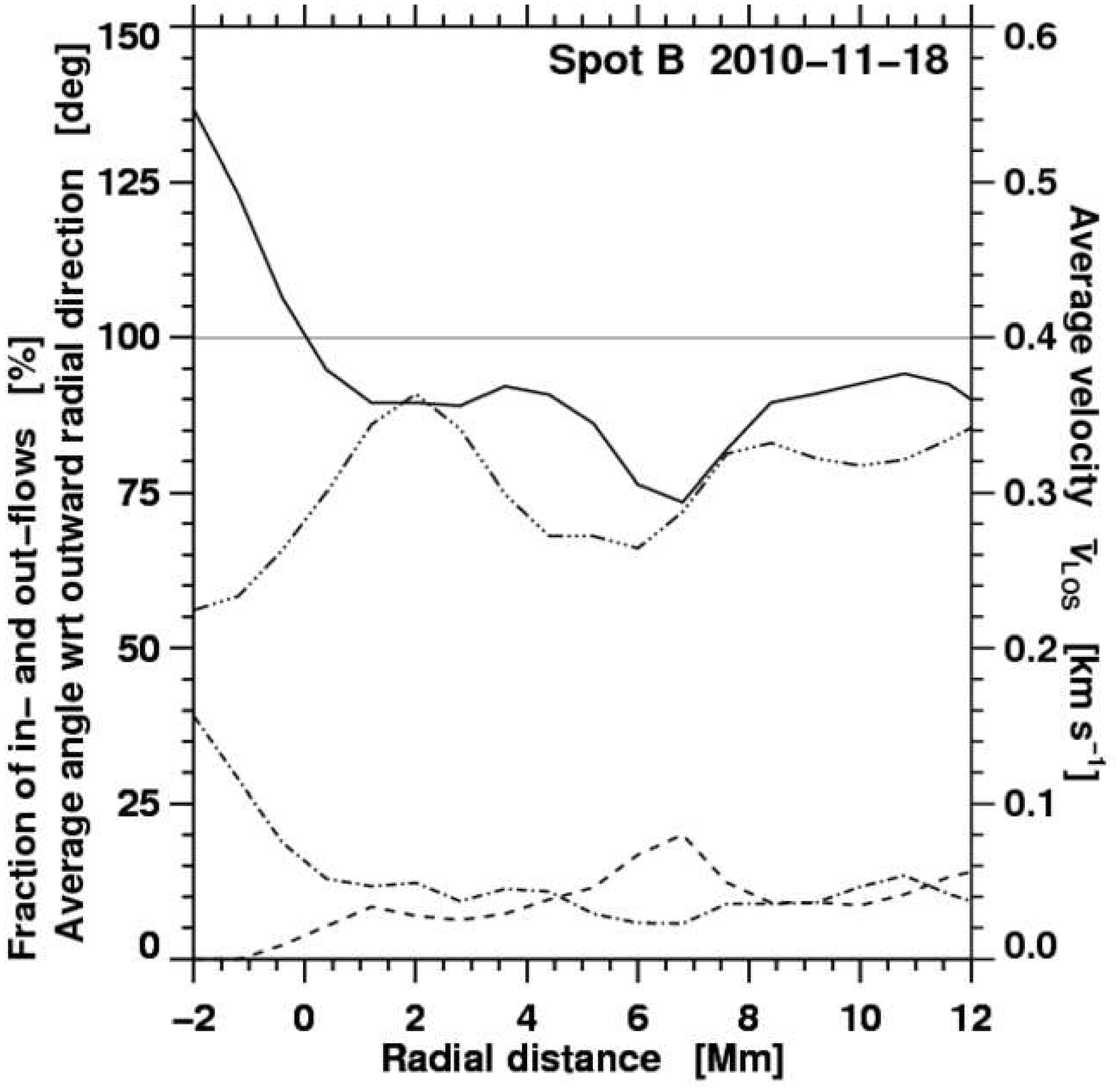}
\caption{Radial dependence of parameters characterizing the flow fields of
    spots~\textsf{A} (\textit{left}) and \textsf{B} (\textit{right}). Radial
    distances are measured from the outer boundary of the spots, i.e., 0.0~Mm
    marks the transition from penumbra to granulation. The thick black curves
    indicate the average angle of horizontal flow vectors with respect to the
    outward radial direction. High values refer to inward flows and low values
    indicate outward flow vectors. Fractions of in-/outward flow vectors
    as defined in the text
    are shown as a dash-dotted and dashed lines, respectively. The horizontal
    straight lines denotes the 100\%-level. The average flow speed is displayed
    as a dash-dot-dot-dotted line.}
\label{FIG07}
\end{figure*}

\subsection{Chromospheric evolution\label{SEC03.2}}

\begin{figure*}[t]
\includegraphics[width=\textwidth]{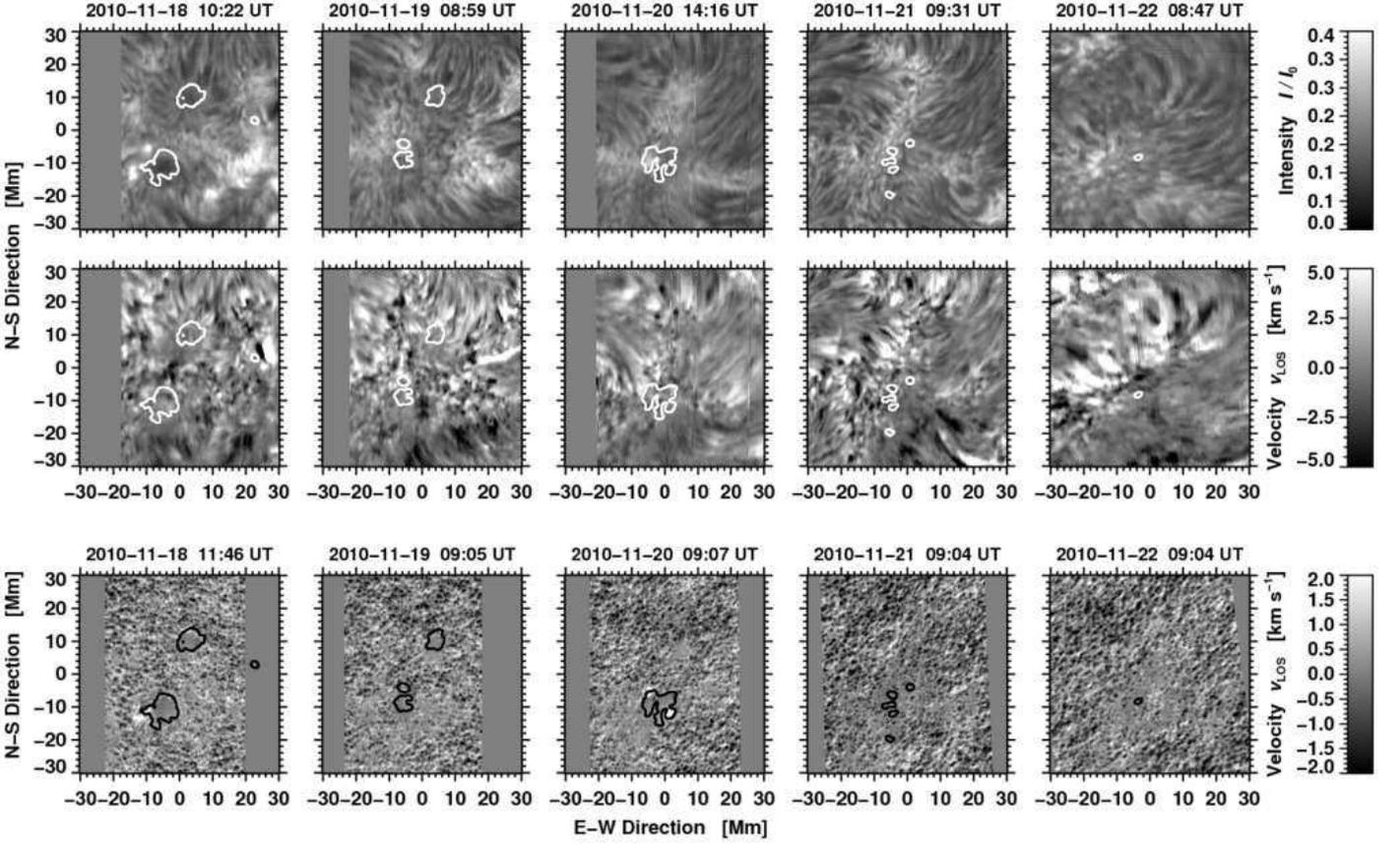}
\caption{VTT Echelle spectra were used to study the
    chromospheric evolution of active region NOAA~11126 during the period from
    2010 November~18--22. The chromospheric fibril structure surrounding the
    decaying sunspots can be traced in H$\alpha$ $\lambda656.28$~nm line core
    intensity (\textit{top}) and LOS velocity (\textit{middle}) maps. In
    addition, photospheric LOS velocities (\textit{bottom}) were derived
    from \textit{Hinode}/SP Fe\,\textsc{i} $\lambda630.25$~nm spectra. The FOV
    matches those of Figs.~\ref{FIG05} and \ref{FIG06}. Regions not covered by
    the spectral data are shown in medium gray. The times above the panels
    refers to the start of a spectral scan.}
\label{FIG08}
\end{figure*}

The description of the chromospheric evolution is based on Ca\,\textsc{ii}\,H
(top row of Fig.~\ref{FIG06}) and H$\alpha$ line core intensity maps (top row of
Fig.~\ref{FIG08}). The Ca\,\textsc{ii}\,H images were averaged over three hours
to highlight some of the long-lived chromospheric features. On November~18, the
lower quarter of the Ca\,\textsc{ii}\,H image showed the undisturbed pattern of
inverse granulation \citep{Rutten2004}. Ca\,\textsc{ii}\,H brightenings cover a
larger area, since their filling factor is significantly greater than that of
G-band bright points. Spot~\textsf{A} was encircled by individual brightenings
at a radial distance of 5~Mm. These individual brightenings coalesced into a
wagon-wheel-like Ca\,\textsc{ii}\,H intensity structure in the three-hour
average image pointing to the presence of MMFs. A similar feature was noticeable
around a much smaller pore with a diameter of about 2~Mm located to the west of
spot~\textsf{A}. This conspicuous Ca\,\textsc{ii}
intensity structure was first described by \citet{Shine1996}, who identified the
bright ridges with azimuthal convergence regions, whereas the dark regions
between the spokes correspond to azimuthal divergence zones. Spot~\textsf{A}
resides in the middle of a supergranular cell with a diameter of about 20~Mm. In
contrast, spot~\textsf{B} is embedded in an area of much more pronounced
Ca\,\textsc{ii}\,H brightenings. The exterior tips of its penumbral filaments
appeared bright because a low-level B-class flare occurred at the time of
observations. 

On November~19, the average Ca\,\textsc{ii}\,H image
showed evidence that the moat flow around spot~\textsf{A} survived the initial
stages of sunspot decay, even after its penumbra had disappeared \citep[see
e.g.,][]{Deng2007}. Spot~\textsf{A} had completely dissolved by November~20.
Assuming that spot~\textsf{A} was located in the center of a supergranular
cell as indicated by the surrounding Ca\,\textsc{ii}\,H brightenings on
November~18 and 19, we conclude that this supergranular cell ceased to exist on
November~20, when the strong magnetic fields of spot~\textsf{A} where no longer
present. The Ca\,\textsc{ii}\,H brightenings are now squeezed together by two
supergranular cells to east and to the west, which could already be identified
on November~19 and remained visible until November~22.

At the time when spot~\textsf{A} vanished, spot~\textsf{B} had already
fragmented into two umbral cores, which were separated by a strong light-bridge
with a noticeable dark lane along its axis \citep{RouppevanderVoort2010}. This
dark core was even more clearly discernible in the three-hour average
Ca\,\textsc{ii}\,H image. The presence of strong
light-bridge can be taken as a first indication of the spot entering the decay
phase. During the next two days on November~21--22, spot~\textsf{B} decayed
further.

The H$\alpha$ line core intensity map of November~18  exposed a radial pattern
of fibrils around spot~\textsf{A} reminiscent of a superpenumbra. Strong
brightenings in the H$\alpha$ line
core in the vicinity of spot~\textsf{B} are related to the aforementioned small
B-class flare. We did not notice any significant  H$\alpha$ intensity features
on the following days, except that after the decay of both spots the H$\alpha$
plage became more prominent. By the end of our observations with the Echelle
spectrograph, only a very compact plage region with a length of about 8~Mm
remained within ROI.

\subsection{Horizontal proper motions\label{SEC03.3}}

The LCT flow maps displayed in Figs.~\ref{FIG05} and~\ref{FIG06} were computed
using the time-sequence of \textit{Hinode} G-band and Ca\,\textsc{ii}\,H images.
As mentioned in Section~\ref{SEC03.1}, we distinguished various solar features
(e.g., G-band bright points, granulation, strong magnetic features, and sunspot
penumbrae) by morphological and adaptive thresholding. For comparison, we refer
to the typical flow speeds of granulation $v_\mathrm{gran} = 0.47 \pm
0.27$~km~s$^{-1}$ and G-band bright points $v_\mathrm{bp} = 0.23 \pm
0.15$~km~s$^{-1}$ as reported by \citet{Verma2011}, who also presented values
for longer time intervals $\Delta T$ (Tab.~1 therein) and cadences $\Delta t$
(Fig.~3 therein). Their values for granulation are in very good agreement with
the mean values $\bar{v}_\mathrm{gran} = 0.36 \pm 0.20$~km~s$^{-1}$ of the
present study. The mean values for bright points
$\bar{v}_\mathrm{bp} = 0.26 \pm 0.15$~km~s$^{-1}$ in the neighborhood of active
region NOAA~11126 are within the range of expected values. Daily values of
flow speeds for granulation and G-band bright points are given in
Tab.~\ref{TAB01}. The standard deviation of the
aforementioned horizontal flow speeds refers to the variance in the data for a
particular solar feature rather than to a formal error. The intrinsic error of
the LCT algorithm is 50~m~s$^{-1}$ and 15$^{\circ}$ for flow speed and
direction, respectively \citep[see][]{Verma2011}. 

We computed the flow vectors for strong magnetic elements, i.e.,
spots~\textsf{A} and \textsf{B}. We identified spots by smoothing the
geometrically corrected G-band images using a Gaussian kernel with a FWHM of
1280~km and subsequently applied an intensity threshold of 0.83$I_{0}$. We
plotted contours  based on these binary masks in Figs.~\ref{FIG05}
and~\ref{FIG06} to provide some visual guidance in identifying the spot
positions in the physical maps. The mean flow speeds $\bar{v}_\mathrm{Spot\ A} =
0.19 \pm 0.17$~km~s$^{-1}$ and $\bar{v}_\mathrm{Spot\ B} = 0.18 \pm
0.11$~km~s$^{-1}$ are virtually identical for both spots. While computing the
mean velocity $\bar{v}_\mathrm{Spot B}$ for spot~\textsf{B}, we discarded
November~22, because on that day only a tiny pore with a diameter of 4~Mm was
left at the location of spot~\textsf{B}. Therefore, adjacent regions with G-band
bright points and granulation would bias the flow speed towards higher
values.The overall flow pattern for the Ca\,\textsc{ii}\,H data is very similar
as compared to G-band data, and the average values $\bar{v}_\mathrm{Spot\ A} =
0.18 \pm 0.20$~km~s$^{-1}$ and $\bar{v}_\mathrm{Spot\ B} = 0.18 \pm
0.08$~km~s$^{-1}$ are virtually identical with a tendency to be slightly lower
on individual days.

The moat flow around sunspots and pores reveals itself as
radially, outward-directed vectors in flow maps, which point to a ring-like
structure with kernels of elevated flow speeds. In the speed and azimuth maps on
November~18 and 19 (Figs.~\ref{FIG05} and~\ref{FIG06}) indications of moat flow
were visible around spot~\textsf{A} and a small pore located to the west of
spot~\textsf{A}. There was no clear signature of the moat flow detectable around
spot~\textsf{B}. To express the characteristics of the moat flow in more
quantitative terms, we plotted in Fig.~\ref{FIG07} the angle of the flow
vectors, the fraction of in- and outward flow vectors, and the horizontal
velocities for spots~\textsf{A} and \textsf{B}. These quantities are radial
averages and the zero point of the radial distance was placed at the
penumbra/granulation boundary of the spots. We considered
a flow vector to point in-/outward, if the tip of the arrow aims in-/outward and
if the angle with the radial direction is less than $20^{\circ}$. Therefore, the
fractions of in-/outward flow vectors presented in Fig.~\ref{FIG07} do not add
up to 100\%, since more tangential flows are neglected. At a radial distance of
$6 \pm 2$~Mm from the boundary of spot~\textsf{A} more than 50\% of flow vectors
point outward. The average flow speed in this region is about 0.5~km~s$^{-1}$.
The region with high-speed outward flows marks the location, where the moat flow
around spot~\textsf{A} terminates. An exact determination of where the moat flow
starts and where it ends strongly depends on the underlying criteria, e.g.,
photometric or magnetic features, or horizontal proper motions.
In Sect.~\ref{SEC03.5}, we will discuss some properties of
moat flow based on the proper motions of MMFs. \citet{Sobotka2007} determine
that the moat radius is independent of the spot size. However,
\citet{Balthasar2010} find that the moat flow terminates at a distance of four
times the spot radius -- in contrast to three times the spot radius in the
present study. Extended statistical studies will help to clarify this issue.

We created two ring-like structures by morphological dilation with a width of
4~Mm, which encircled both spots at a distance of 2~Mm. This ring-like structure
starts, where inflows turn to outflows 2~Mm beyond the boundary of
spot~\textsf{A}, and terminates, where the outflows reach the highest speed. We
used these templates to calculate the flows in the immediate neighborhood of
spots~\textsf{A} and \textsf{B}. These regions were labeled as rings~\textsf{A}
and \textsf{B}. The velocities for these regions are included in
Tab.~\ref{TAB01}. The flow speed $v_\mathrm{Ring\ A}$ was more than double than
that of ${v}_\mathrm{Spot\ A}$ because of the suppressed convective motions in
the spot's interior. The flow speeds in the immediate vicinity of
spot~\textsf{B} was higher by 30--40\% on November~18 and 19. However, on
November~22 during the final decay stage of spot~\textsf{B}, no major difference
between the spot and its closest surroundings were observed.

To effectively visualize the plasma motions we presented high resolution maps of
flow speed and azimuth (see Figs.~\ref{FIG05} and~\ref{FIG06}). The speed maps
of G-band and Ca\,\textsc{ii}\,H images are virtually same. The region with
G-band bright points exhibits suppressed velocities in both cases. The regions
around spot~\textsf{A} and the small pore to the west are surrounded by a ring
of high velocities (around 0.5~km~s$^{-1}$), where the ordered moat flow
terminates. In the azimuth maps of November 18 and 19 outward plasma motions are
traceable in these regions. In addition, to gain more insight into inflows and
outflows around the spots we computed the divergence of the flow field. The
divergence maps are included in the bottom row of Fig.~\ref{FIG05}. The positive
values of divergence refer to outflows, whereas negative values indicate
inflows. Negative divergence values are encountered within the boundaries of the
sunspots. In contrast small patches of positive divergence encircle the spots.
This is indicative of inflows in the sunspots and outflows at their periphery.
The area with low divergence values is increasing as the active region is
decaying. The presence of a low divergence region could be a signature of the
final stages of decay.

\begin{figure*}[t]
\includegraphics[width=\textwidth]{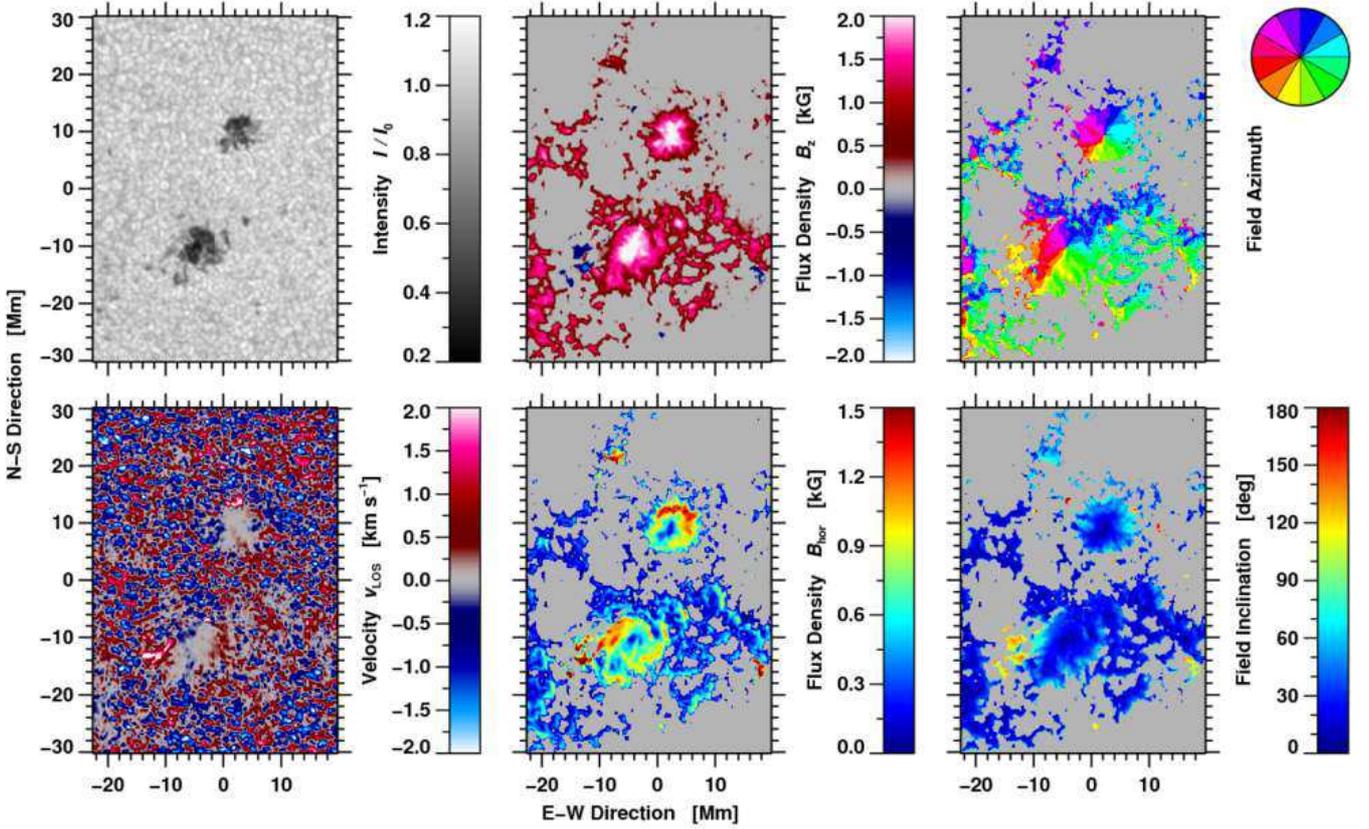}
\caption{Maps of physical parameters derived using SIR-code for
    \textit{Hinode}/SP scan on 2010 November~18 (\textit{from top-left to
    bottom-right}) normalized intensity $I / I_0$, vertical component of
    magnetic flux density $B_\mathrm{z}$, magnetic field azimuth $\phi$, Doppler
    velocity $v_\mathrm{LOS}$, horizontal component of magnetic flux density
    $B_\mathrm{hor}$, and magnetic field inclination $\gamma$.}
\label{FIG09}
\end{figure*}

\subsection{Line-of-sight velocities\label{SEC03.4}}

The LOS velocities were derived using \textit{Hinode}/SP \ion{Fe}{i} spectral
data and H$\alpha$ spectra of the VTT Echelle spectrograph. LOS velocities for
the \ion{Fe}{i} spectral line were calculated using the Fourier phase method
\citep{Schmidt1999}. This method makes use of entire line profile and is less
sensitive to noise, along with this it takes into account the spectral line
asymmetry. To compute the LOS velocity for H$\alpha$ spectra we calculated
shifts using parabola fits to the central 50~pixels (0.03~nm), because the
H$\alpha$ spectral line is too wide to identify the real continuum. The
calculated shifts in both cases were converted to velocities using the Doppler
formula. The average photospheric velocity of dark umbral cores  was used as the
frame of reference. The Fe\,\textsc{i} $\lambda$656.92~nm line served as the
reference for the H$\alpha$ velocities. The Doppler velocity maps are displayed
in the bottom and middle rows of Fig.~\ref{FIG08}. Redshifts in these maps are
positive and blueshifts are negative, hence, areas moving away from the observer
are bright, while areas moving toward the observer are dark.

To compute velocities for various solar features we used an intensity mask
generated using G-band images. The values are compiled in Tab.~\ref{TAB01}. On
November~18 and 20 in the Fe\,\textsc{i} Doppler velocity
maps strong photospheric downflows were observed at edges of spot~\textsf{B}.
The average
downflow velocity in these regions is about 2.5~km~s$^{-1}$. In all maps a gray
patch of near zero velocity was observed in the central FOV, which became more
prominent on the last two days of observation. We identify this region with
abnormal granulation \citep[see e.g.,][]{DeBoer1992}, in which convection is
still strongly inhibited by the presence of (dispersed) magnetic fields.

In case of H$\alpha$ LOS velocities no conspicuous features were visible in the
velocity maps, except on November~18 when we observed a radial filamentary
structure around spot~\textsf{A} in H$\alpha$ line core intensity maps, which
resembled a superpenumbra. At the footpoints of the dark H$\alpha$ filaments, we
measure downflows from about 3.5 up to 4.5~km~s$^{-1}$, which we interpret as an
inverse Evershed flow \citep{Maltby1975}. Tightly wound superpenumbral spirals
are only predicted for spots with radii larger than 8~Mm, whereas
spot~\textsf{A} was compact with a radius of 4~Mm. The downward chromospheric
velocities at the edge of spot~\textsf{A} are compatible with the MHD model of
superpenumbral flows presented by \citet{Peter1996}.


\subsection{Magnetic fields\label{SEC03.5}}

In addition to the photometric evolution shown in Fig.~\ref{FIG03}, we also
computed the flux contained in the active region as shown in Fig.~\ref{FIG04}.
Since only HMI LOS magnetograms were available (processing of the vector
magnetograms is still under way). We took the measured
magnetic field strength at face value and only carried out a geometrical
correction to yield the proper average values of the magnetic flux. The
geometrical correction only applies to the surface area covered by a pixel,
which can simply be achieved by dividing the magnetic field strenght by $\mu=cos
\theta$. Signatures of geometric projection effects can be seen when the active
region was close to the east limb, and the angle between LOS and shallow
penumbral field lines leads to an apparent polarity reversal. Consequently, the
positive and negative flux gradients have opposite signs, while to total flux
remained almost constant for the first 30~hours until projection effects become
less severe. To compute the temporal evolution of magnetic flux, we created a
binary template, which only contained pixels above/below $\pm 50$ Gauss in the
HMI magnetograms. Morphological erosion with a 1-Mm kernel was applied to the
template three times to eliminate small, isolated flux concentrations. Finally,
we used morphological dilation with a 5-Mm kernel to include the strong magnetic
fields in the immediate neighborhood of active region NOAA~11126. In this way,
we avoided a bias, which could be introduced by weak magnetic fields, which do
not belong to the active region but are contained within the FOV shown in
Fig~\ref{FIG02}.

The growth rates of the magnetic flux are 2.66, 1.98, and $1.09 \times
10^{13}$~Wb day$^{-1}$ for the total, positive, and negative flux, respectively.
The negative flux showed a monotonous rise until about 06:00~UT on November~16,
whereas the positive flux increased with a twice steeper slope, stopped a day
earlier, and turned to a shallower slope until the end of November~17. On
average the positive flux was three times larger than the negative one. However,
using adaptive thresholding and morphological image processing tools, we only
measured the flux in the immediate vicinity of the sunspots. The missing
negative flux required for flux balance has to be contained in flux
concentration beyond the immediate neighborhood of the sunspots. The decay rates
of the magnetic flux are four to five times lower than the growth rates and
amount to 0.66, 0.47, and $0.23 \times 10^{13}$~Wb day$^{-1}$ for the total,
positive, and negative flux, respectively. A linear fit to the data was used and
despite some deviations from a linear trend, there is no indication for a
parabolic (or any other) decay law. The decay rates of the present study are in
agreement with \citet{Kubo2008a} who report rates on the order of $10^{13}$~Wb
day$^{-1}$ and discusses the magnetic flux loss rate in a decaying active
region.

The SIR-code (Stokes Inversion based on Response functions) developed by
\citet{RuizCobo1992} was used to invert the \textit{Hinode}/SP spectra. We
restrict ourselves to the more sensitive line Fe\,\textsc{i}\
$\lambda$630.25\,nm ($g_\mathrm{eff} = 2.5$). The starting model covers the
optical depth range $+1.0 \le \log \tau \le -4.4$. A limb-darkening factor is
considered according to Eq.\,10 of \citet{Pierce1977b}. We assume a constant
macroturbulence of 1\,km\,s$^{-1}$ and a fixed straylight contribution of one
percent. The inversions deliver the temperature stratification with four nodes
$T(\tau)$, the total magnetic flux density $B_\mathrm{tot}$, the magnetic
inclination $\gamma$ and azimuth $\phi$, and the Doppler velocity
$v_\mathrm{LOS}$ constant with height (one node for each of these physical
parameters).

The magnetic azimuth ambiguity must be solved after the inversions. For the
first two days when we observed two major pores/small sunspots, we assumed two
azimuth centers \citep[see][]{Balthasar2006}. For the other days it is
sufficient to assume one azimuth center away from the pores in the direction
towards disk center. The magnetic field in such small pores is more or less
vertical to the solar surface so that on these days the field is sufficiently
inclined with respect to the LOS that such an assumption is justified. If the
expected azimuth $\phi$ deviates by more than 90$^{\circ}$ and less than
270$^{\circ}$ from the calculated one, we correct it by 180$^{\circ}$. Finally,
we rotate the magnetic vector with respect to the local solar frame. For a few
locations, it happens that there is a sudden change of sign in the Cartesian
components of the magnetic vector. This problem is solved by an additional
correction of the LOS-azimuth. To be on the safe side, we additionally apply to
data of November~18 the code provided by \citet{Leka2009}, which minimizes
$\vert J_z\vert + \vert \nabla B\vert$ to solve the azimuth ambiguity. Pixels,
where the integrated circular and the integrated linear polarization are below
0.006, are excluded from the further analysis of the magnetic field.
We finally use the magnetic vector field in the local
solar frame. The results of the SIR-inversion for November~18 are depicted in
Fig.~\ref{FIG09}.

The vertical magnetic flux density $B_z$ in the ROI is predominantly positive
and points outwards. The only significant patches of negative polarity were
found on November~18 on the eastern side of spot~\textsf{B}. Penumbral filaments
connect spot~\textsf{B} to several magnetic knots of negative polarity. In
addition, we measure Evershed flows with velocities $v_\mathrm{LOS} >
2$~km~s$^{-1}$. On smaller spatial scales, we find MMFs in the vicinity of
spot~\textsf{A}. They are mostly of type~II (unipolar with the same polarity as
the spot) but a few scattered \textsf{U}-shaped type~I MMFs (bipolar with the
inner footpoint of opposite polarity of the spot) were observed as well. No
type~I MMFs were observed near spot~\textsf{B}.

\begin{figure*}[t]
\sidecaption
\includegraphics[width=0.66\textwidth]{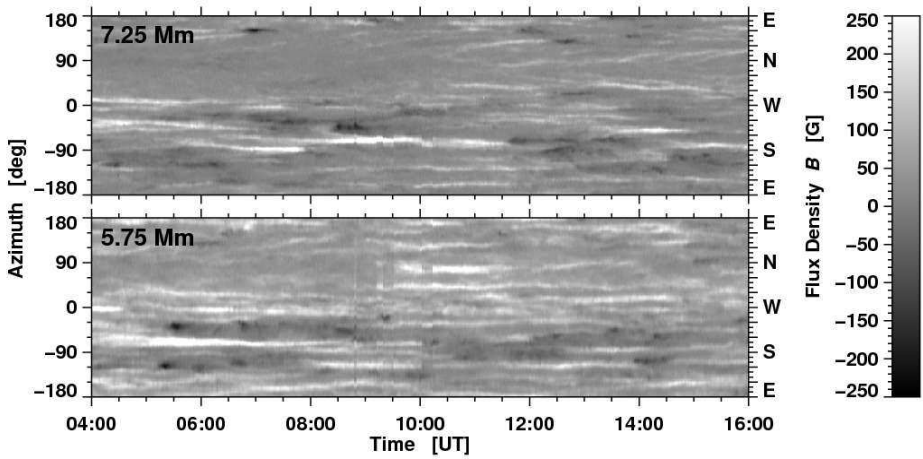}
\caption{Space-time slices showing the temporal evolution of the magnetic flux
    along annuli with radii of 5.75~Mm (\textit{bottom}) and 7.25~Mm
    (\textit{top}), respectively, which are centered on spot~\textsf{A}. HMI
    magnetograms with a cadence of 45~s covered a period of 12~hours starting at
    04:00~UT on 2010 November~18. The annulus with a widths of 0.5~Mm was cut
    open in the East and then transformed to straight line so that the southern
    part of the spot is mapped to the lower half of the space-time map. The
    heliographic direction is indicated on the axis to the right.}
\label{FIG10}
\end{figure*}

Since the time cadence of \textit{Hinode}/SP scans is
about 12~min, we used SDO/HMI magnetograms with a cadence of 45~s to study the
dynamics of MMFs. Space-time slices are an option to visualize local changes of
the magnetic flux in a time-series. The continuous coverage of HMI magnetograms
allowed us to depict in Fig.~\ref{FIG10} the temporal evolution of the magnetic
flux around spot~\textsf{A} during a 12-hour period on 2010 November~18. The
magnetic flux changes were recorded within two 0.5-Mm wide annuli of 5.75~Mm
(bottom panel) and 7.25~nm (top panel) radius, respectively. Note that the
linear scale of the the ordinate in Fig.~\ref{FIG10} is 38~Mm and 47~Mm for the
inner and outer ring, respectively. The inner ring corresponds to the location,
where the strongest outflows were observed (see Fig.~\ref{FIG07}). The outer
ring marks the site, where the largest fraction (about 8\%) of the annulus is
covered by negative-polarity features.

\begin{figure}[t]
\centerline{\includegraphics[width=\columnwidth]{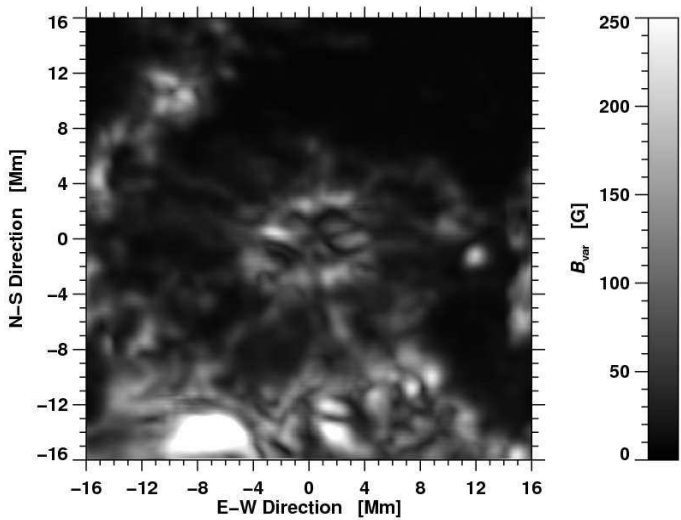}}
\caption{Local, background-subtracted variation of the magnetic flux
    $B_{\mathrm{var}}$ around spot~\textsf{A} on 2010 November 18.}
\label{FIG11}
\end{figure}

Within a radial distance from 4--10~Mm from the center
of spot~\textsf{A}, only about 5\% of the magnetic flux observed during
12-hour period is of negative polarity. Negative-polarity MMFs are weaker by
about 20\% as compared to MMFs of opposite polarity. Negative flux elements are
most prominent in the southern half of spot~\textsf{A} and absent in the
northern half. The most notable difference between positive- and
negative-polarity features is the morphology. Negative-polarity features can be
tracked in space-time slices for only about 15--30~min, i.e., they are strongly
localized. Time-lapse movies show that these negative-polarity features are
mostly type~I MMFs. The positive polarity appears first in the space-time slices
(to the left) followed by the negative polarity. The appearance of the
positive-polarity features is very different. Horizontal striation can be seen
for several hours. They can abruptly (dis-)appear and in a few cases merging as
well as splitting of the striae can be observed. In time-lapse movies, type~II
MMFs are moving away from spot~\textsf{A} along radial paths on top of a
background of positive flux.

The temporal variation of the magnetic flux above/below the local background is
given by $B_{\mathrm{var}} = \langle | B - \langle B \rangle | \rangle$, where
$\langle \ldots \rangle$ indicates a temporal average. To compute the temporal
average in Fig.~\ref{FIG11}, we used the same 12-hour time-series of HMI
magnetograms that was used to create Fig.~\ref{FIG10}. The strongest changes
occur at the periphery of spot~\textsf{A}, where also the horizontal magnetic
field $B_{\mathrm{hor}}$ is strongest as can be seen in Fig.~\ref{FIG09}. This
correspondence also applies to the feature of strong horizontal field to the
north-east of spot~\textsf{A}, which belongs to the (chromospheric) network. The
most notable features in Fig.~\ref{FIG11} are, however, several spoke-like
structures with a length of about 10~Mm, which extend from spot~\textsf{A} all
the way out to the border of the supergranule surrounding the spot.
These elongated structures are the signatures of MMFs
streaming from the spot along identical paths \citep{Harvey1973} towards the
supergranular boundary. However, the distance traveled is much shorter as
\citet{Harvey1973} observed for mature sunspots. Therefore, we conclude that the
preferred paths of MMFs are a phenomenon, which can be observed even in the
final stages of sunspot decay. The traces left by MMFs in the variance of the
magnetic field around its background $B_{\mathrm{var}}$ are most prominent on
the eastern side of the spot. They also match the wagon-wheel structure
discussed in the context of the time-averaged Ca\,\textsc{ii}\,H images in
Sect.~\ref{SEC03.2}. Finally, the preference of MMFs for distinct radial
channels connecting the sunspots' magnetic fields to the strong network fields
was also noticed by \citet{Hagenaar2005}.

Maps similar to Fig.~\ref{FIG09} were computed for each
day so that we could study the evolution of the magnetic field. The strongest
horizontal magnetic flux densities $B_\mathrm{hor}$ are observed in the
immediate surroundings of spots~\textsf{A} and \textsf{B} as long as the spots
are compact (November~18 and 19). They symmetrically enclose the entire boundary
of the spots. Once spot~\textsf{A} had disappeared and spot~\textsf{B} started
to fragment, this symmetry is broken most evidently on November~20, when strong
horizontal fields were present at the western side of spot~\textsf{B}. This
corresponds to a time when a rudimentary penumbra was present. Whenever
indications of penumbral filaments were observed during the decay of the
sunspots, inclined magnetic field lines and Evershed flows were present at the
same time \citep[cf.,][]{Leka1998, Yang2003a}.

The magnetic field lines spread out symmetrically from the center of the spots
on  November~18 and 19. Such well defined azimuth centers can still be found in
the later decay stages of spot~\textsf{B}. Even though these region are no
longer circular and become elongated. Smaller azimuth centers are also observed
to the east of spot~\textsf{B} on November~19. These centers are associated to
several magnetic knots.

The Doppler velocity $v_{\mathrm{LOS}}$ is suppressed in the presence of strong
magnetic features. Starting on November~20, the velocity pattern associated with
the magnetic fields changes. Positive Doppler velocities occupy more and more of
the magnetic region. This coherent pattern was also observed in the divergence
maps (bottom row of Fig.~\ref{FIG05}) and the horizontal flow maps
(Figs.~\ref{FIG05}  and \ref{FIG06}). Indeed, considering that the region is
approaching the limb, some of the Doppler velocities can be interpreted as a
coherent proper motion of the magnetic region towards the south-west (cf. the
azimuth maps of the horizontal proper motions). In summary, during the final
stages of sunspot decay, the three-dimensional flow field, in regions previously
occupied by strong magnetic fields, significantly differs from granular flow
patterns or regions of reduced velocities in the presence of G-band bright
points. Since we are only presenting a case study, the question remains, if this
intriguing flow pattern is a typical feature of sunspot decay.


\section{Conclusions\label{SEC04}}

We have presented a detailed account of the final stages in the decay of the
active region NOAA~11126,  which did not obey the Hale-Nicholson polarity law
\citep[e.g.,][]{Zirin1988a}.  Since only one out of ten active regions shows
such a behavior \citep{Howard1990} and we only present a case study, our results
might not be representative for sunspot decay in general. However, space
missions such as \textit{Hinode} and SDO provide nowadays data of sufficient
coverage, resolution, and cadence that statistical properties of sunspot decay
become accessible. Furthermore, previous studies of non-Hale regions
\citep[e.g.,][and reference therein]{LopezFuentes2000} were centered on flux
emergence, $\delta$-spots, and strong solar flares. The present study can
consequently be considered as an extension of these studies with a focus on a
much quieter magnetic field topology, which might be representative for the
lower solar activity during cycle No. 24 \citep{Petrovay2010, Nielsen2011}.

The major findings of our study can be summarized as follows: (1) MMFs were
observed in the vicinity of spot~\textsf{A} until it decayed. Mostly type~II and
a few interspersed \textsf{U}-shaped type~I MMFs contributed to the observed
moat flow, which also left a clear signature in the time-averaged
Ca\,\textsc{ii}\,H images \citep{MartinezPillet2002}. (2) Even though penumbral
filaments had almost completely disappeared in photospheric G-band images of
spot~\textsf{A} on November~18, H$\alpha$ line core images clearly exhibited a
structure reminiscent of a superpenumbra. Thus, filamentary structures including
the inverse Evershed flow \citep{Maltby1975, Georgakilas2003} might be more
persistent in the chromosphere. (3) We have also observed MMFs in the vicinity
of a tiny pore with a diameter of about 2~Mm, which did not show any indication
of penumbral filaments. Such an observation argues strongly against a close tie
between Evershed flows and MMFs \citep[cf., ][]{VargasDominguez2008,
VargasDominguez2010, CabreraSolana2006}. \citet{Rempel2011} argues based on MHD
simulations of sunspots that penumbral flows can be separated in two components,
where the shallow one corresponds to the Evershed flow and the deeper one is
related to moat flow. (4) The strong rotation and twist seen in spot~\textsf{B}
might explain, why this trailing spot never advected sufficient magnetic flux to
establish more than a rudimentary penumbra and remained highly fragmented during
its entire life cycle. (5) The photospheric and chromospheric maps of horizontal
flows show a peculiar pattern, once the last dark feature of the active region
has disappeared. In general, the flow field in this region is less structured
than regions covered by granulation, i.e., the dispersed magnetic field still
significantly affects the convective pattern. Chromospheric flows have increased
notably compared to times when spots and pores were still present. Most
prominently, a contiguous area of low divergence appears towards the end of
sunspot decay.


\begin{acknowledgements}
\textit{Hinode} is a Japanese mission developed and launched by ISAS/JAXA,
collaborating with NAOJ as a domestic partner, NASA and STFC (UK) as
international partners. Scientific operation of the \textit{Hinode} mission is
conducted by the \textit{Hinode} science team organized at ISAS/JAXA. This team
mainly consists of scientists from institutes in the partner countries. Support
for the post-launch operation is provided by JAXA and NAOJ (Japan), STFC (UK),
NASA, ESA, and NSC (Norway). The SDO/HMI images are provided by the Joint
Science Operations Center (JSOC) Science Data Processing (SDP). The Vacuum Tower
Telescope at the Spanish Observtorio del Teide of the Instituto de
Astrof\'{\i}sica de Canarias is operated by the German consortium of the
Kiepenheuer-Institut f\"ur Sonnenphysik in Freiburg, the Leibniz-Institut f\"ur
Astrophysik Potsdam, and the Max-Planck-Institut f\"ur Sonnensystemforschung in
Katlenburg-Lindau. MV expresses her gratitude for the generous financial support
by the German Academic Exchange Service (DAAD) in the form of a PhD scholarship.
CD was supported by grant DE~787/3-1 of the German Science Foundation (DFG). CL
and HW were supported by NSF grants AGS~08-19662 and AGS~08-49453, and NASA
grants NNX~08AQ90G and NNX~08AJ23G. ND was supported by NASA grant NNX~08AQ32G.
The authors would like to thank Drs.\ Alexandra Tritschler and Klaus G.\
Puschmann as well as the referee Dr.\ Luis Bellot Rubio for
carefully reading the manuscript and providing ideas, which significantly
enhanced the paper.
\end{acknowledgements}



\end{document}